\documentstyle[wstwocl,axodraw]{article}
\begin{document}

\bibliographystyle{unsrt}    % for BibTeX - sorted numerical labels

\newenvironment{comment}[1]{}{}
\newcommand{\tenrm}{\mbox{}}
\newcommand{\st}{\scriptstyle}
\newcommand{\sst}{\scriptscriptstyle}
\newcommand{\mco}{\multicolumn}
\newcommand{\epp}{\epsilon^{\prime}}
\newcommand{\vep}{\varepsilon}
\newcommand{\ra}{\rightarrow}
\newcommand{\ppg}{\pi^+\pi^-\gamma}
\newcommand{\vp}{{\bf p}}
\newcommand{\ko}{K^0}
\newcommand{\kb}{\bar{K^0}}
\newcommand{\al}{\alpha}
\newcommand{\ab}{\bar{\alpha}}
\def\be{\begin{equation}}
\def\ee{\end{equation}}
\def\bea{\begin{eqnarray}}
\def\eea{\end{eqnarray}}
\def\CPbar{\hbox{{\rm CP}\hskip-1.80em{/}}}%temp replacement due to no font

\def\ap#1#2#3   {{\em Ann. Phys. (NY)} {\bf#1} (#2) #3.}
\def\apj#1#2#3  {{\em Astrophys. J.} {\bf#1} (#2) #3.}
\def\apjl#1#2#3 {{\em Astrophys. J. Lett.} {\bf#1} (#2) #3.}
\def\app#1#2#3  {{\em Acta. Phys. Pol.} {\bf#1} (#2) #3.}
\def\ar#1#2#3   {{\em Ann. Rev. Nucl. Part. Sci.} {\bf#1} (#2) #3.}
\def\cpc#1#2#3  {{\em Computer Phys. Comm.} {\bf#1} (#2) #3.}
\def\err#1#2#3  {{\it Erratum} {\bf#1} (#2) #3.}
\def\ib#1#2#3   {{\it ibid.} {\bf#1} (#2) #3.}
\def\jmp#1#2#3  {{\em J. Math. Phys.} {\bf#1} (#2) #3.}
\def\ijmp#1#2#3 {{\em Int. J. Mod. Phys.} {\bf#1} (#2) #3.}
\def\jetp#1#2#3 {{\em JETP Lett.} {\bf#1} (#2) #3.}
\def\jpg#1#2#3  {{\em J. Phys. G.} {\bf#1} (#2) #3.}
\def\mpl#1#2#3  {{\em Mod. Phys. Lett.} {\bf#1} (#2) #3.}
\def\nat#1#2#3  {{\em Nature (London)} {\bf#1} (#2) #3.}
\def\nc#1#2#3   {{\em Nuovo Cim.} {\bf#1} (#2) #3.}
\def\nim#1#2#3  {{\em Nucl. Instr. Meth.} {\bf#1} (#2) #3.}
\def\np#1#2#3   {{\em Nucl. Phys.} {\bf#1} (#2) #3.}
\def\pcps#1#2#3 {{\em Proc. Cam. Phil. Soc.} {\bf#1} (#2) #3.}
\def\pl#1#2#3   {{\em Phys. Lett.} {\bf#1} (#2) #3.}
\def\prep#1#2#3 {{\em Phys. Rep.} {\bf#1} (#2) #3.}
\def\prev#1#2#3 {{\em Phys. Rev.} {\bf#1} (#2) #3.}
\def\prl#1#2#3  {{\em Phys. Rev. Lett.} {\bf#1} (#2) #3.}
\def\prs#1#2#3  {{\em Proc. Roy. Soc.} {\bf#1} (#2) #3.}
\def\ptp#1#2#3  {{\em Prog. Th. Phys.} {\bf#1} (#2) #3.}
\def\ps#1#2#3   {{\em Physica Scripta} {\bf#1} (#2) #3.}
\def\rmp#1#2#3  {{\em Rev. Mod. Phys.} {\bf#1} (#2) #3.}
\def\rpp#1#2#3  {{\em Rep. Prog. Phys.} {\bf#1} (#2) #3.}
\def\sjnp#1#2#3 {{\em Sov. J. Nucl. Phys.} {\bf#1} (#2) #3.}
\def\spj#1#2#3  {{\em Sov. Phys. JEPT} {\bf#1} (#2) #3.}
\def\spu#1#2#3  {{\em Sov. Phys.-Usp.} {\bf#1} (#2) #3.}
\def\zp#1#2#3   {{\em Zeit. Phys.} {\bf#1} (#2) #3.}

\setcounter{secnumdepth}{2} % Number sections and subsections

%%%%%%%%%%%%%%%%%%%%%%%%%%%%%%%%%%%%%%%%%%%%%%%%%%
%                                                %
%    BEGINNING OF TEXT                           %
%                                                %
%%%%%%%%%%%%%%%%%%%%%%%%%%%%%%%%%%%%%%%%%%%%%%%%%%

%\title{\LARGE\bf Supersymmetric Searches in $e^-e^-$, $e^-\gamma$ and
%%$\gamma\gamma$ Scattering}
\title{\LARGE\bf Slepton Production in Polarized \boldmath$\gamma\gamma$
Scattering}

\firstauthors{Frank Cuypers}

\firstaddress{{\tt cuypers@mppmu.mpg.de}\\
        Max-Planck-Institut f\"ur Physik,
        Werner-Heisenberg-Institut,
        F\"ohringer Ring 6,
        D--80805 M\"unchen,
        Germany}

\bigskip

\twocolumn[\faketitle
%\abstracts{}
]

In this talk
I have summarized the potential of
supersymmetric searches in $e^-e^-$, $e^-\gamma$ and $\gamma\gamma$ scattering.
Detailed studies have already been published
for $e^-e^-$ and $e^-\gamma$ collisions\cite{ee,eg}.
I therefore do not repeat these results here,
but rather report an as yet unpublished analysis
of slepton pair-production in polarized $\gamma\gamma$ scattering.
It is an extension of what was presented in Ref.\cite{gg},
where discovery limits were derived
for unpolarized backscattered laser photon beams.
Here I show that the combined use
of polarization and angular correlations,
can greatly enhance the resolving power of the
$\gamma\gamma$ reaction.
Chargino pair-production in polarized $\gamma\gamma$ scattering
has also been studied recently\cite{kon}.

\begin{figure}[htb]
%\vskip2mm
\begin{center}
{\unitlength.6mm
\SetScale{1.703}%\SetScale{1.419}%\SetScale{2.837}
\begin{picture}(50,20)(-10,0)
\Photon(0,0)(15,10){-1}{3.5}
\Text(-1,0)[r]{\normalsize$\gamma$}
\Photon(0,20)(15,10){1}{3.5}
\Text(-1,20)[r]{\normalsize$\gamma$}
\DashLine(15,10)(30,20){1}
\Text(31,20)[l]{\normalsize$\tilde \ell^+$}
\DashLine(15,10)(30,0){1}
\Text(31,0)[l]{\normalsize$\tilde \ell^-$}
\end{picture}}
\qquad
{\unitlength.6mm
\SetScale{1.703}%\SetScale{1.419}%\SetScale{2.837}
\begin{picture}(50,20)(0,0)
\Photon(0,0)(15,0){1}{3}
\Text(-1,0)[r]{\normalsize$\gamma$}
\Photon(0,20)(15,20){-1}{3}
\Text(-1,20)[r]{\normalsize$\gamma$}
\DashLine(15,20)(15,0){1}
\Text(16,10)[l]{\normalsize$\tilde \ell$}
\DashLine(15,20)(30,20){1}
\Text(31,20)[l]{\normalsize$\tilde \ell^+$}
\DashLine(15,0)(30,0){1}
\Text(31,0)[l]{\normalsize$\tilde \ell^-$}
\end{picture}}
\raisebox{5mm}{+ crossed}
\end{center}
\caption{Feynman diagrams.}
\vskip-2mm
\label{feyn}
\end{figure}
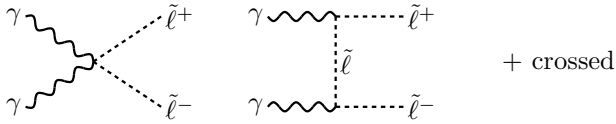

The lowest order Feynman diagrams
for slepton pair-production
are shown in Fig.~\ref{feyn}.
It is followed by the decays of the sleptons
into their corresponding leptons and neutralinos.
In the framework of the minimal supersymmetric standard model
the latter escape detection.
The signal is thus
an acoplanar lepton pair and missing energy.
Obviously,
this analysis applies as well to selectrons as to smuons,
and to a less clean extent to staus.

The differential and total cross sections for this process are
\begin{eqnarray}
\hskip-5mm & \hskip-7mm \displaystyle{\mbox{d}\sigma\over\mbox{d}t} =
\displaystyle{\pi\alpha^2\over s} \hskip-2mm &
\Biggl\{
  1 + \left( 1 + 2{m^2\over t-m^2} + 2{m^2\over u-m^2} \right)^2
  \nonumber\\ &&
  + ~2~P_1P_2~ \left( 1 + 2{m^2\over t-m^2} + 2{m^2\over u-m^2} \right)
\Biggr\}
\label{xsd}
\\
\hskip-5mm & \hskip-7mm \sigma = \displaystyle{\pi\alpha^2\over s} \hskip-3mm &
\Biggl\{
  (1+x) \sqrt{1-x} - x \left(1-{x\over2}\right)
\ln{1+\sqrt{1-x}\over1-\sqrt{1-x}}
  \nonumber\\ &&
  + ~P_1P_2~ \left( \sqrt{1-x} - x \ln{1+\sqrt{1-x}\over1-\sqrt{1-x}} \right)
\Biggr\}
\label{xst}
\end{eqnarray}
where $s,t,u$ are the Mandelstam variables,
$m$ is the slepton mass,
$x=4m^2/s$ and
$P_{1,2}$ are the polarizations of the photon beams.

At a linear collider of the next generation
operating in the TeV range,
%and obeying the luminosity scaling law
%\begin{equation}
%${\cal L} = 80 s_{ee}/\mbox{TeV}^2$
%\label{lum}~
%\end{equation}
%for electron beams.
%At such a machine
hard and intense photon beams can be obtained
by backscattering an intense laser on the incoming electron beams\cite{gin}.
The cross sections (\ref{xsd},\ref{xst}) must thus be folded up
with the resulting photon spectra.
There is also an unavoidable loss of luminosity
which is taken into account here as in Ref.\cite{eg}.
I also assume 100\%\ polarized lasers
and 95\%\ polarized electron beams.

To warrant a high photon luminosity
one must live with\cite{gin}
$\sqrt{s_{\gamma\gamma}} \le .83\sqrt{s_{ee}}$.
Therefore this collider mode is in general not suited for discoveries.
However,
because the initial state photons only couple to the charge
of the produced particles,
there is also little room left for any model dependence
and the $\gamma\gamma$ mode is thus ideally suited
for studying their properties.
This is also the case for supersymmetric studies:
by the time the $\gamma\gamma$ mode is turned on,
the sleptons would have already been discovered
and their masses determined
in either $e^-e^-$ or $e^+e^-$ collisions;
the $\gamma\gamma \to \tilde\ell^+\tilde\ell^-$ cross section
would then be accurately predicted
and the measurement of the
$\gamma\gamma \to \ell^+\ell^- + E_{\rm mis}$
cross section
would provide a direct determination
of the
$\tilde\ell \to \ell \tilde\chi^0_1$
branching ratio.
The situation is rendered slightly more complicated
by possible cascade decays of the sleptons\cite{casc},
but these small corrections
can only be taken seriously into account
with a complete detector simulation.

\begin{figure}[htb]
\begin{center}
\input{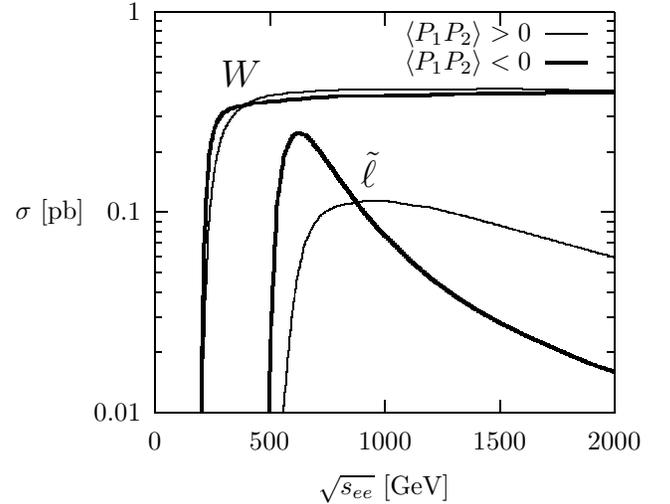}
\end{center}
\vskip-5mm
\caption{
  Energy dependence of a 200 GeV slepton signal and
  its $W$ background uncut cross sections.
}
\vskip-2mm
\label{eny}
\end{figure}

The only background which cannot be rendered harmless
by a simple cut on the $p_\perp$ of the emerging leptons,
originates from $W^+W^-$ pair-production
and subsequent leptonic decay\cite{gg}.
The energy dependence
of the cross sections
of this background and the signal of a 200 GeV slepton
are displayed in Fig.~\ref{eny}.
Clearly,
the best signal to background ratio is obtained
in the threshold region
when the two photon beams have opposite average polarizations.
It turns out that
the highest cross sections are obtained
when the electron beams' centre of mass energy
is approximately three times the slepton mass.
This remains true for any reasonable slepton mass.

\begin{comment}{
In Fig.~\ref{mas},
I have plotted the energy dependence
of the signal cross section
for several slepton masses
and for photon beams with opposite average polarizations.
Assuming the $W$ pair-production background can be handled
(which is indeed the case)
%{\em cf.} Fig.~\ref{cont2})
even sleptons as heavy as 800 GeV
would still be produced with sufficient rates
at a 2 TeV machine
to allow studying their properties
over a large portion of the supersymmetry parameter space.

\begin{figure}[htb]
\begin{center}
\input{mas.tex}
\end{center}
\vskip-5mm
\caption{Feynman diagrams.}
\vskip-5mm
\label{mas}
\end{figure}
}\end{comment}

The signal to background ratio can be further enhanced
by using the information
undoubtedly gathered in previous $e^-e^-$ or $e^+e^-$ experiments:
the energies of the observed leptons
which originate from the supersymmetric signal
are confined between
\begin{equation}
        E_\ell \in {E\over4} \left[1-{m^2_{\tilde\chi^0_1}\over
m^2_{\tilde\ell}}\right]
                \left[1\pm\sqrt{1-{4m^2_{\tilde\ell}\over E^2}}\right]
\label{elep}~,
\end{equation}
where $E=.83\sqrt{s_{ee}}$
is the maximum attainable $\gamma\gamma$ centre of mass energy.
Imposing this cut on the data
significantly reduces the $W^+W^-$ background
at no cost for the supersymmetric signal.

\begin{figure}[htb]
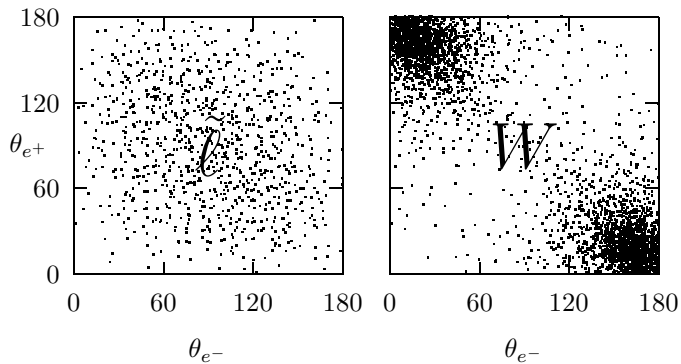

\hskip-17mm
\input{scats.tex}
\hskip-25mm
\input{scatw.tex}
\vskip-5mm
\caption{
  Angular distributions of the events
  from a 300 GeV slepton signal and
  its $W$ background.
  The collider's $e^\pm e^-$ energy and luminosity are 1 TeV and 80 fb$^{-1}$.
  No cuts are applied.
}
\vskip-2mm
\label{scat}
\end{figure}

Furthermore,
as can be gathered from Fig.~\ref{scat},
the supersymmetric leptonic events
are rather uniformly distributed at all angles
with a slight dominance in the transverse direction.
On the other hand,
the leptons which originate from the decays of the $W$'s
are mostly emitted at small angles from the beampipe.

\begin{figure}[htb]
\begin{center}
\input{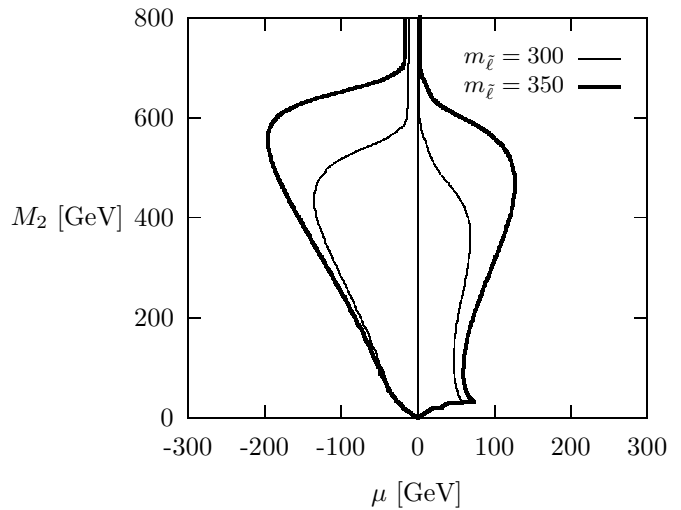}
\end{center}
\vskip-5mm
\caption{
  Contours of observability of two slepton signals
  for $\tan\beta=4$.
  The collider's $e^\pm e^-$ energy and luminosity are 1 TeV and 80 fb$^{-1}$.
  The cuts (\protect\ref{elep},\protect\ref{cuts}) are applied.
}
\vskip-2mm
\label{cont1}
\end{figure}

To estimate the resolving power of the $\gamma\gamma$ mode,
one can therefore divide the $(\theta_{e^-},\theta_{e^+})$ angular range
into $3\times3$ equal size two-dimensional bins
and perform a least squares test
%and compute the least squares function
%\begin{equation}
%\chi^2 = \displaystyle\sum_i^{3\times3}
%\left(
%        {n_i^{\rm exp}-n_i^{\rm SM} \over \Delta n}
%\right)^2
%~\geq~6~
%\label{chi2}~
%\end{equation}
for different values of the relevant supersymmetry parameters.
Demanding $\chi^2\ge6$,
one obtains in Fig.~\ref{cont1}
contours in the $(\mu,M_2)$ plane
of supersymmetry parameters
for 300 and 350 GeV sleptons
produced at a 1 TeV collider
with 80 fb$^{-1}$ of $e^\pm e^-$ luminosity.
The following cuts have been implemented
in addition to Eq.~(\ref{elep}):
\begin{equation}
10^o < \theta_{e^\pm} < 170^o
\quad~
p^\perp_{e^\pm} > 10 \mbox{ GeV}
\quad~
\phi_{\mbox{acopl}} > 2^o
\label{cuts}~.
\end{equation}
These cuts very efficiently remove the potential backgrounds
from the standard model lepton pair-production mechanisms
and their electroweak Bremsstrahlung\cite{gg}.

As it turns out,
the two contours in Fig.~\ref{cont1}
delimit very precisely the regions of parameter space
where the branching ratio
$\tilde\ell \to \ell \tilde\chi^0_1$
is .02 or .035
for the sleptons of mass 300 and 350 GeV respectively.
The branching ratios become smaller
inside the contours.
Clearly,
this direct measurement of the slepton branching ratios
will provide invaluable information about the nature of supersymmetry breaking.

%\begin{figure}[htb]
%\begin{center}
%\input{gg2000.tex}
%\end{center}
%\vskip-5mm
%\caption{Feynman diagrams.}
%\vskip-5mm
%\label{cont2}
%\end{figure}

To conclude,
the $\gamma\gamma$ operating mode
of a linear collider of the next generation
can be advantageously used
to constrain the parameter space of supersymmetry
by providing an accurate measurement of the branching ratio
of sleptons into their leptonic partners and the lightest neutralino.

\setcounter{secnumdepth}{0} %this ensures that there are no section numbers
                            %from here on in the text. Don't remove.

\end{document}